\documentclass[preprint,amsmath,amssymb,superscriptaddress,floatfix]{revtex4-2}
\usepackage[utf8]{inputenc}
\usepackage[english]{babel}
\DeclareUnicodeCharacter{202A}{}
\usepackage{graphicx}
\usepackage{makecell}
\usepackage{color}
\usepackage{hyperref}
\usepackage{lineno}
\usepackage{filecontents}

\begin{document}

% \linenumbers

\title{Lattice-distortion couplings in antiferroelectric perovskite $\rm AgNbO_3$ and comparison with $\rm PbZrO_3$}
\author{Huazhang Zhang}
\email[Corresponding author: ]{hzhang@uliege.be}
\address{Theoretical Materials Physics, Q-MAT, Université de Liège, B-4000 Sart-Tilman, Belgium}
\address{Department of Physics, School of Science, Wuhan University of Technology, Wuhan 430070, People’s Republic of China}
\author{Konstantin Shapovalov}
\address{Theoretical Materials Physics, Q-MAT, Université de Liège, B-4000 Sart-Tilman, Belgium}
\author{Safari Amisi}
\address{Theoretical Materials Physics, Q-MAT, Université de Liège, B-4000 Sart-Tilman, Belgium}
\address{Laboratoire de Physique des Solides et des Interfaces, Institut Supérieur Pédagogique de Bukavu, Democratic Republic of the Congo}
\author{Philippe Ghosez}
\email[Corresponding author: ]{philippe.ghosez@uliege.be}
\address{Theoretical Materials Physics, Q-MAT, Université de Liège, B-4000 Sart-Tilman, Belgium}
\date{\today}

\begin{abstract}
Lead-free antiferroelectric perovskite $\rm AgNbO_3$ is nowadays attracting extensive research interests due to its promising applications in energy storage. 
Although great progress has been made in optimizing the material performance, fundamental questions remain regarding the mechanism stabilizing the antiferroelectric $Pbcm$ phase.
Here, combining structural symmetry analysis and first-principles calculations, we identified crucial anharmonic couplings of oxygen octahedra rotations and cation antipolar motions which contribute significantly to lowering the energy of the $Pbcm$ phase.
The stabilization of this phase shows close similarities with the stabilization of the $Pbam$ phase in $\rm PbZrO_3$ except that in $\rm AgNbO_3$ the octahedra rotations are the primary distortions while the antipolar cation motions appear to be secondary. The appearance and significant amplitude of the latter are explained from the combination of hybrid-improper and triggered mechanisms.

\begin{description}
\item[Keywords]
Antiferroelectricity; silver niobate; lead zirconate; lattice-distortion coupling
\end{description}
\end{abstract}

\maketitle
\newpage
\section{Introduction}

Antiferroelectric (AFE) materials exhibit characteristic double hysteresis loops, making them attractive for a wide range of applications \cite{RN724, Rabe2013, RN408, RN532, RN526, RN531}.
Among the limited number of perovskite antiferroelectric oxides, $\rm AgNbO_3$ (ANO) has attracted a lot of research interest because it is lead-free and can exhibit large polarization under electric field (52 $\rm \mu C/cm^2$ at 220 $\rm kV/cm$ \cite{RN315}).
Recently, extensive studies have focused on the energy storage properties of ANO-based materials and remarkably high energy storage density has been achieved \cite{RN457, RN316, RN631, RN318, RN309, RN319}.
The excellent energy storage performance of ANO is closely related to its AFE property, which involves a reversible phase transition between nonpolar AFE state and a highly polarized ferroelectric (FE) state upon application and removal of an electric field.
Understanding the mechanism behind the AFE properties of ANO and identifying the key interactions stabilizing the AFE state are fundamental and meaningful topics, which can provide valuable guidance for the further optimization of its properties.

The exact structure and the AFE nature of ANO are currently subjects of ongoing discussion.
In fact, controversies exist regarding its complex series of phase transitions \cite{RN314, RN348, RN310, RN352, RN356, RN324}.
According to the X-ray and neutron diffraction reported by Sciau {\it et al.} \cite{RN352}, the phase sequence of ANO from high-temperature to low-temperature is: cubic $C$ phase ($Pm\bar{3}m$, above $852$ K), tetragonal $T$ phase ($P4/mbm$, $660 \sim 852$ K), orthorhombic $O$ phase ($Cmcm$, $626 \sim 660$ K), and orthorhombic $M$ phase ($Pbcm$, below $626$ K).
Although dielectric measurements suggested that there may be several different $M$ phases \cite{RN314, RN348, RN319}, Sciau's results showed that the average structures of all these possible low-temperature $M$ phases share the same $Pbcm$ space group \cite{RN352}.
The $Pbcm$ structure is centrosymmetric, in line with its AFE property.
However, it was also noted that ANO exhibits a weak remanence on the polarization hysteresis loops \cite{RN357, RN319}.
This phenomenon may be caused by various reasons, such as cation disordering, polar clusters, or defects \cite{RN356, RN327, RN319}.
A previous structural study also suggested that a polar $Pmc2_1$ structure might be the low-temperature phase of ANO \cite{RN325}, but the subsequent density functional theory (DFT) calculations have shown that this $Pmc2_1$ structure is not stable \cite{RN445}.
Instead, the $Pbcm$ phase remains the lowest energy phase in literature and is free of any dynamical instability, including under negative pressure \cite{RN445}, indicating that it is the most probable antiferroelectric phase of ANO.

The emergence of a $Pbcm$ phase is rather uncommon in perovskite oxides \cite{Howard08,Amisi23}. 
The most common structure for perovskites with small tolerance factors ($t < 1$) is $Pnma$ \cite{RN598}, a phase which is favored by a trilinear coupling between cation antipolar motions and in-phase and antiphase oxygen octahedra rotations \cite{RN44}.
Then, among antiferroelectrics, the phase that attracted the most attention is $Pbam$, as seen in, e.g., the prototypical antiferroelectric $\rm PbZrO_3$ (PZO), where it is stabilized by a trilinear coupling between cation displacements, oxygen octahedra rotations and their modulations \cite{Iniguez2014, Hlinka2014,amisi2021ab, ShapovalovandStengel}.
In ANO, however, the stabilization mechanism of the $Pbcm$ antiferroelectric state has not been fully clarified.

In this study, we investigate the stabilization mechanism of the $Pbcm$ antiferroelectric state through structural symmetry analysis and first-principles calculations, aiming at understanding why such an unusual phase in perovskites can exhibit a low energy in ANO. 
We identify the combination of trilinear and cooperative biquadratic couplings between cation antipolar motions and oxygen octahedra rotations as the key contributions to the energy lowering of the $Pbcm$ phase.
We also conduct a comparison between ANO and the prototypical antiferroelectric PZO.
Our findings suggest a combination of hybrid-improper and triggered mechanisms for the appearance of the cation antipolar motions in ANO.

\section{Calculation details}
First-principles DFT calculations were performed using the \textsc{Abinit} package \cite{Gonze2020, Gonze2016, Gonze2009, Gonze2002} with the plane-wave-pseudopotential approach.
We employed the generalized gradient approximation (GGA) with the revised Perdew-Burke-Enzerh parameterization for solids (PBEsol) \cite{PBEsol} and optimized norm-conserving pseudopotentials from the PseudoDojo server  \cite{VANSETTEN201839, Hamann2013}.
The valence electron configurations are $4s^2 4p^6 4d^{10} 5s^1$ for Ag, $4s^2 4p^6 4d^4 5s^1$ for Nb, $5s^2 5p^6 5d^{10} 6s^2 6p^2$ for Pb, $4s^2 4p^6 4d^2 5s^2$ for Zr, and $2s^2 2p^4$ for O.
For comparison, we also checked some results using the local density approximation (LDA) functional \cite{LDA}.
The energy cutoff for the plane-wave expansion was 50 $\rm Ha$ for ANO and 60 $\rm Ha$ for PZO.
The $k$-point grids were equivalent to or denser than $6 \times 6 \times 6$ grid for the five-atom cubic perovskite.
The electronic self-consistent cycles were considered to be converged until the potential residual is smaller than $10^{-18}$ $\rm Ha$. 
Structural relaxations were performed based on the Broyden-Fletcher-Goldfarb-Shanno minimization algorithm, using the convergence criteria of $10^{-6}$ $\rm Ha/Bohr$ for the forces and $10^{-8}$ $\rm Ha/Bohr^3$ for the stresses.
The phonon dispersions were calculated according to density functional perturbation theory as implemented in \textsc{Abinit} and analyzed with the \textsc{Anaddb} program \cite{DFPT}.
The symmetry-adapted lattice-distortion modes were analyzed using \textsc{ISODISTORT} \cite{ISODISTORT}.
The energy expansions were build with the help of \textsc{INVARIANTS} \cite{INVARIANTS}.

\section{Mode couplings in ANO}
The structural parameters of ANO in its $Pbcm$ phase are reported in Table \ref{tab:lattparam}, where we compare the results of our GGA-PBEsol and LDA calculations with the experimental data from low-temperature measurements in literature \cite{RN352}. 
The GGA-PBEsol calculated lattice parameters show very good agreement with the experimental values, with discrepancies less than 0.5\%, while the LDA tends to underestimate the lattice constants. 
It is therefore expected that the GGA-PBEsol functional is more accurate in capturing the structural characteristics of ANO.

\begin{table*}[ht]
\caption{Lattice parameters (\AA) and atomic positions (reduced coordinates) of ANO in the orthorhombic $Pbcm$ phase. The $x$, $y$, $z$ coordinates are in the direction of orthorhombic axes $a$, $b$, $c$, respectively.}
\label{tab:lattparam}
\begin{ruledtabular}
\begin{tabular}{cccccccccc}
             &\multicolumn{3}{c}{GGA-PBEsol}&\multicolumn{3}{c}{LDA}&\multicolumn{3}{c}{Exp. (1.5 K) \cite{RN352}}\\
\colrule
Lattice      &  $a$    &  $b$    &  $c$    &  $a$    &  $b$    &  $c$    &  $a$    &  $b$    & $c$     \\
parameter (\AA)   &  5.5215 &  5.6019 &  15.506 &  5.4424 &  5.5362 &  15.390 &  5.5436 &  5.6071 &  15.565 \\
Error (\%)   & -0.40   & -0.09   & -0.38   & -1.83   & -1.26   & -1.12   & \\
\colrule
             &  $x$    &  $y$    &  $z$    &  $x$    &  $y$    &  $z$    &  $x$    &  $y$    &  $z$    \\
Ag1 (4c)     &  0.7423 &  0.25   &  0      &  0.7439 &  0.25   &  0      &  0.7449 &  0.25   &  0      \\
Ag2 (4d)     &  0.7369 &  0.2210 &  0.25   &  0.7378 &  0.2226 &  0.25   &  0.7420 &  0.2211 &  0.25   \\
Nb  (8e)     &  0.2436 &  0.2210 &  0.1250 &  0.2445 &  0.2270 &  0.1250 &  0.2436 &  0.2212 &  0.1249 \\
O1  (8e)     & -0.0438 &  0.0445 &  0.1082 & -0.0473 &  0.0479 &  0.1067 & -0.0348 &  0.0371 &  0.1098 \\
O2  (8e)     &  0.5445 &  0.4620 &  0.1417 &  0.5462 &  0.4603 &  0.1436 &  0.5360 &  0.4689 &  0.1396 \\
O3  (4c)     &  0.3167 &  0.25   &  0      &  0.3234 &  0.25   &  0      &  0.3109 &  0.25   &  0      \\
O4  (4d)     &  0.1872 &  0.2752 &  0.25   &  0.1794 &  0.2787 &  0.25   &  0.1944 &  0.2678 &  0.25   \\
\end{tabular}
\end{ruledtabular}
\end{table*}

\begin{figure*}[ht]
\centering
\includegraphics[scale=0.45]{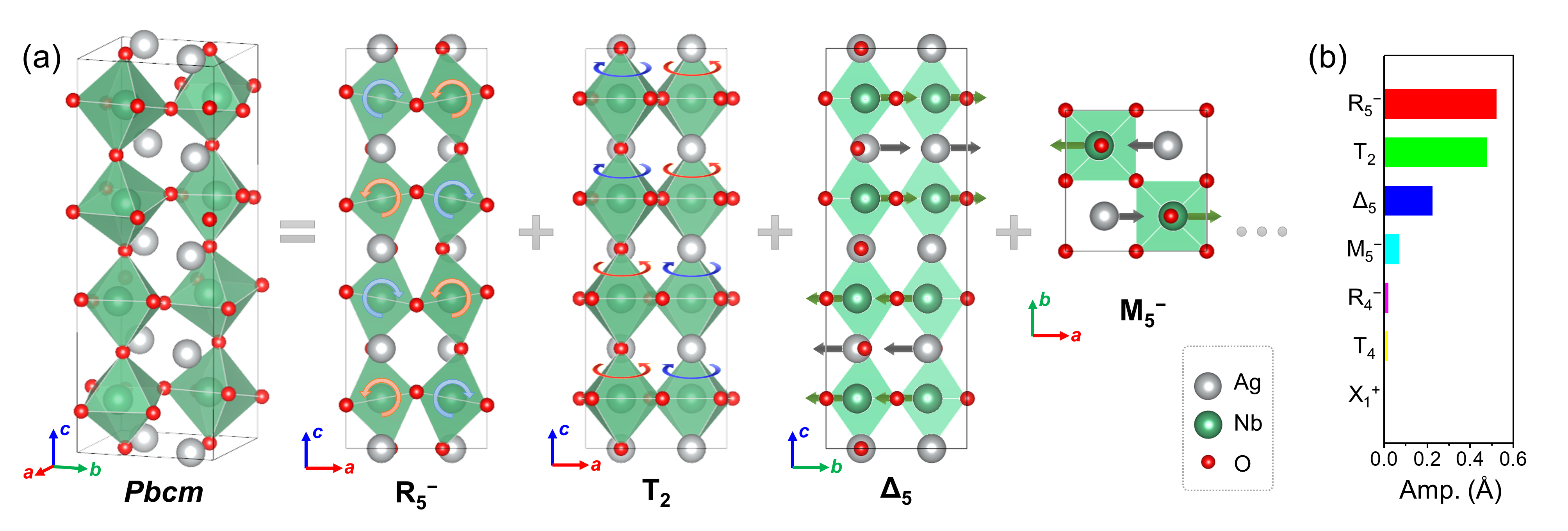}
\caption{Structure and symmetry-adapted lattice-distortion modes (respect to the cubic reference structure) of ANO in the $Pbcm$ phase. (a) Schematics of the $Pbcm$ crystal structure and the dominant modes of the cubic reference structure giving rise to the $Pbcm$ phase. 
(b) Mode amplitudes in the $Pbcm$ structure, respect to the cubic reference, as optimized with the GGA-PBEsol functional.}
\label{fig:struct}
\end{figure*}

As shown in Fig. \ref{fig:struct}(a), the unit cell of the $Pbcm$ phase can be viewed as a $\sqrt{2} \times \sqrt{2} \times 4$ multiple of the five-atom perovskite elemental unit cell, with the translational vectors $a$, $b$, $c$ directed along the pseudocubic $[1\bar10]$, $[110$], $[001]$ directions. 
By symmetry analysis, the atomic distortion of the $Pbcm$ phase with respect to the cubic reference structure can be decomposed into symmetry-adapted lattice-distortion modes of this reference [Fig. \ref{fig:struct}(b) and Table \ref{tab:modeANO}].
The dominating modes are [Fig. \ref{fig:struct}(a)]: 
(1) $R_5^-$ mode, the rotations of oxygen octahedra around the $b$-axis that correspond to $a^-a^-c^0$ rotation pattern in Glazer’s notation \cite{RN722};
(2) $T_2$ mode, a complex octahedra rotation pattern alternating between in-phase and antiphase rotations around the $c$-axis that we label hereafter as the $a^0a^0c^{+/-}$ rotation pattern;
(3) $\Delta_5$ mode, primarily the antipolar motions of the cations Ag and Nb in the direction of $b$-axis;
(4) $M_5^-$ mode, the antipolar motions of Ag and Nb in the direction of $a$-axis.
The above four modes account for 48.88\%, 41.13\%, 9.02\% and 0.86\% of the total lattice distortion, respectively, while the remaining modes account for the remaining 0.11\%.
The fact that most of the distortions are associated to the oxygen octahedra rotations is consistent with the small tolerance factor of ANO ($t = 0.965$) since octahedra rotations are typical in perovskites with tolerance factor smaller than 1 \cite{RN598}.

\begin{table*}[ht]
\caption{Symmetry-adapted lattice-distortion mode decomposition of the $Pbcm$ phase of ANO. 
The structure is obtained by relaxing fully the atomic positions and lattice parameters using GGA-PBEsol functional.
The mode amplitudes are reported in the so-called ``parent-cell-normalized" values of ISODISTORT.}
\label{tab:modeANO}
\begin{ruledtabular}
\begin{tabular}{clllllll}
 &  $R_5^-$  &  $T_2$     & $\Delta_5$ & $M_5^-$   & $R_4^-$   &   $T_4$   & $X_1^+$   \\
\colrule                                                                          
Amplitude (\AA) & $0.5202$ & $0.4772$  & $0.2235$  & $0.0691$ & $0.0177$ & $0.0164$ & $0.0005$ \\
Percentage (\%) & $48.88$   & $41.13$    & $9.02$     & $0.86$    & $0.06$    & $0.05$    & $< 0.01$  \\
\end{tabular}
\end{ruledtabular}
\end{table*}

\begin{figure*}[ht]
\centering
\includegraphics[scale=0.45]{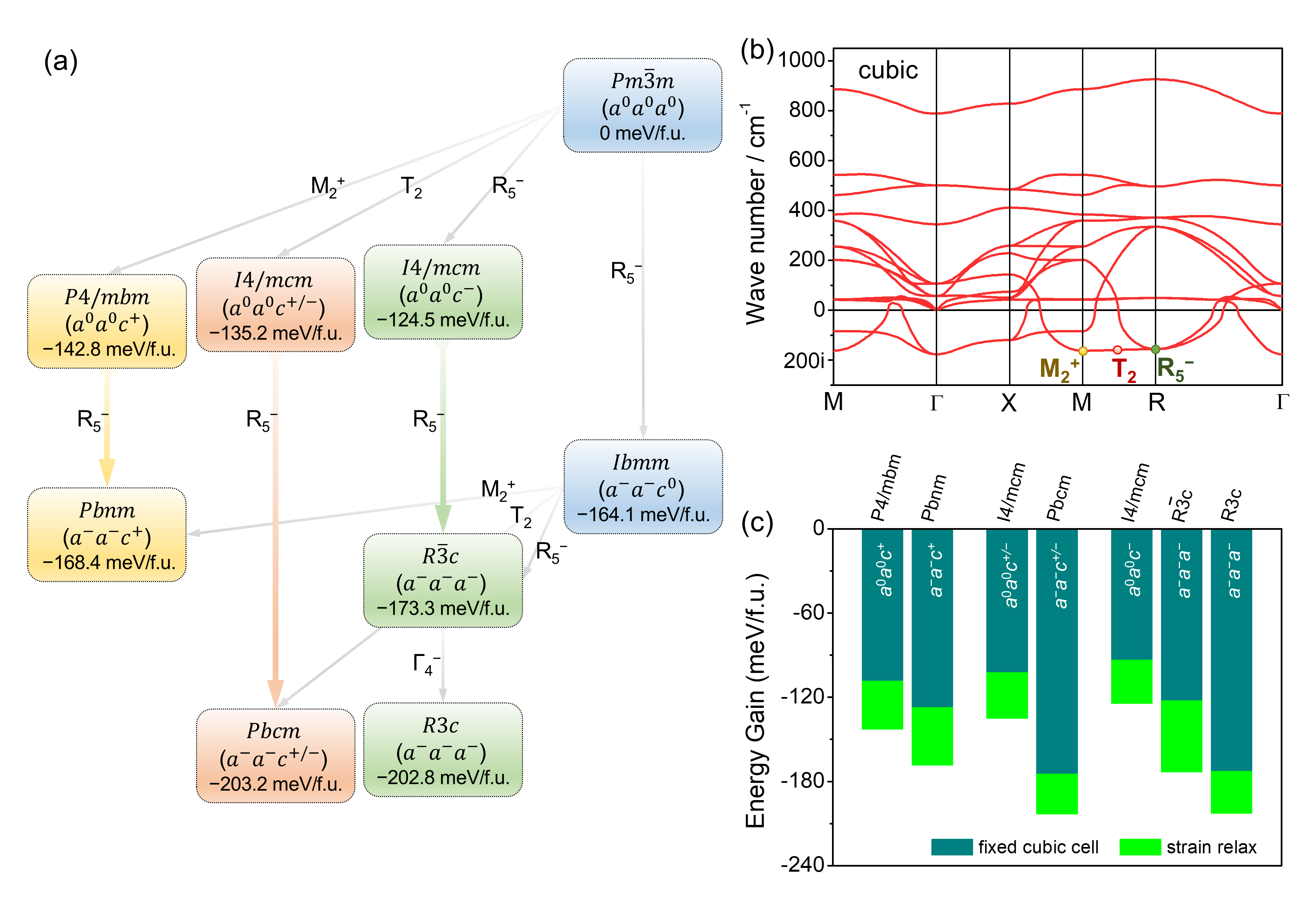}
\caption{(a) Structural and energetic relationships of various metastable phases. 
(b) Phonon dispersion curves of ANO in the cubic parent phase. 
(c) Comparisons of the energies of the metastable phases (respect to the cubic phase taken as reference) with either fixed lattice parameters of the cubic parent phase or fully relaxed lattice parameters.
The notations for orthorhombic space groups $Pbnm$ and $Ibmm$ are chosen to keep consistency with the crystallographic coordinate system of this work, but it should be noted that these space groups are commonly referred to as $Pnma$ and $Imma$ in the literature, respectively, as they are equivalent.}
\label{fig:phases}
\end{figure*}

We can gain more insight into the stabilization mechanism of the $Pbcm$ phase by investigating the structural and energetic relationships among different phases of ANO, paying specific attention to the phases with different oxygen octahedra rotations. 
In each calculation, we introduce octahedra rotations to the prototypical cubic $Pm\bar3m$ phase and relax the atomic distortions and lattice parameters while preserving the symmetry of the distortions in order to obtain the corresponding metastable phase.
Specifically, we consider three types of octahedra rotation patterns around the $c$-axis, namely $a^0a^0c^+$, $a^0a^0c^{+/-}$, and $a^0a^0c^-$, which are corresponding to the $M$ $(1/2, 1/2, 0)$, $T$ $(1/2, 1/2, 1/4)$, and $R$ $(1/2, 1/2, 1/2)$ points in reciprocal space, respectively, as well as the octahedra rotations around $b$-axis corresponding to $a^-a^-c^0$ rotation pattern in Glazer's notation.

Figure \ref{fig:phases}(a) schematically illustrates the structural and energetic relationships among various phases.
When only one rotation around the $c$-axis is present, the three phases $P4/mbm$ ($a^0a^0c^+$), $I4/mcm$ ($a^0a^0c^{+/-}$) and $I4/mcm$ ($a^0a^0c^-$) are very close in energy [Fig. \ref{fig:phases}(a)], which is consistent with the almost non-dispersive nature of the lowest $M-T-R$ branch of the phonon dispersion curve calculated in the cubic phase [Fig. \ref{fig:phases}(b)].
This implies that the rotations of oxygen octahedra, while naturally strongly correlated within the planes perpendicular to the rotation axis $c$, are weakly correlated between adjacent planes.
Consequently, the rotations between adjacent planes being either in-phase or antiphase do not result in significant energy differences.

Interestingly, when the octahedra rotations around the $c$-axis coexist with the $a^-a^-c^0$ rotations, we find that the energies of the relaxed $Pbnm$ ($a^-a^-c^+$, equivalent to $Pnma$ but hereafter will be referred to it as $Pbnm$ to keep consistency with the crystallographic coordinate system), $Pbcm$ ($a^-a^-c^{+/-}$) and $R\bar{3}c$ ($a^-a^-a^-$) phases are no longer so close to each other. 
In particular, the energy lowering from $I4/mcm$ ($a^0a^0c^{+/-}$) to $Pbcm$ ($a^-a^-c^{+/-}$) is notably larger than in the other two cases [i.e. from $P4/mbm$ ($a^0a^0c^+$) to $Pbnm$ ($a^-a^-c^+$), and from $I4/mcm$ ($a^0a^0c^-$) to $R\bar{3}c$ ($a^-a^-a^-$)].
As we will show below, such comparatively larger energy lowering is caused by condensation of additional non-polar modes trilinearly coupled to the octahedra rotations.
In addition to this, $R\bar{3}c$ phase allows a further lowering of its symmetry towards $R3c$ by introducing a polar mode $\Gamma_4^{-}$ along the total rotation axis in the $a^-a^-a^-$ pattern (pseudocubic $[111]$ direction), resulting in a large energy lowering.
Nevertheless, the energy of the $R3c$ phase is still a bit higher than that of the $Pbcm$ phase.

The impact of strain relaxation is examined in Fig. \ref{fig:phases}(c), where, in addition to the above results, we report the energies of the phases obtained at fixed lattice parameters of the cubic parent phase.
It can be seen that fixing the lattice parameters does not qualitatively change the energetics of the phases: the energy reduction from $I4/mcm$ ($a^0a^0c^{+/-}$) to $Pbcm$ ($a^-a^-c^{+/-}$) is as pronounced as with strain relaxation, with the $Pbcm$ phase still slightly below the $R3c$ phase in energy.
This rules out the strain-phonon coupling to be a dominant factor for stabilizing the $Pbcm$ phase.

We then turn to analyze in more details how the lattice-distortion modes are coupled in ANO.
We will mainly focus on the $Pbcm$ phase and take the $Pbnm$ and $R3c$ phases for comparisons.
We will adopt Landau-type energy expansion in terms of symmetry-adapted lattice-distortion modes for the understanding of the mode couplings, and neglect strain relaxation which does not play a dominant role.

For the $Pbcm$ phase, we consider the aforementioned four dominating modes, i.e. $R_5^-$, $T_2$, $\Delta_5$ and $M_5^-$.
Taking these modes as order parameters and restricting to the lowest necessary fourth order, the Landau energy expansion for the $Pbcm$ phase is written as
\begin{equation}
\begin{split}
E_{Pbcm} =
& E_{\rm cubic}
+ A_R Q_R^2 + B_R Q_R^4 
+ A_T Q_T^2 + B_T Q_T^4
+ A_\Delta Q_\Delta^2 + B_\Delta Q_\Delta^4 \\
& + A_{M^\prime} Q_{M^\prime}^2
+ C_{R T \Delta} Q_R Q_T Q_\Delta
+ C_{T \Delta M^\prime} Q_T Q_\Delta Q_{M^\prime} \\
& + D_{R T} Q_R^2 Q_T^2
+ D_{R \Delta} Q_R^2 Q_\Delta^2
+ D_{T \Delta} Q_T^2 Q_\Delta^2,
\end{split}
\label{eq:Pbcm}
\end{equation}
where $E_{\rm cubic}$ is the energy of the cubic reference structure, $Q_R$, $Q_T$, $Q_\Delta$ and $Q_{M^\prime}$ are the amplitudes of the $R_5^-$, $T_2$, $\Delta_5$ and $M_5^-$ modes, respectively, and the $A$, $B$, $C$, $D$ parameters are the coefficients of the energy expression, with subscript indicating the associated modes.
For comparisons, we also write the energy expansions for the $Pbnm$ phase
\begin{equation} 
\begin{aligned}
E_{Pbnm} =
& E_{\rm cubic}
+ A_R Q_R^2 + B_R Q_R^4 
+ A_M Q_M^2 + B_M Q_M^4 
+ A_X Q_X^2 \\
& + C_{R M X} Q_R Q_M Q_X 
+ D_{R M} Q_R^2 Q_M^2
+ D_{R X} Q_R^2 Q_X^2
+ D_{M X} Q_M^2 Q_X^2,
\end{aligned}
\label{eq:Pbnm}
\end{equation}
where $Q_R$, $Q_M$, and $Q_X$ are the amplitudes of the $R_5^-$, $M_2^+$, and $X_5^-$ dominant modes, respectively, and for the $R3c$ phase
\begin{equation} 
\begin{aligned}
E_{R3c} =
& E_{\rm cubic}
+ A_{R^\prime} Q_{R^\prime}^2 + B_{R^\prime} Q_{R^\prime}^4 
+ A_\Gamma Q_\Gamma^2 + B_\Gamma Q_\Gamma^4 
+ D_{R^\prime \Gamma} Q_{R^\prime}^2 Q_\Gamma^2,    
\end{aligned}
\label{eq:R3c}
\end{equation}
where $Q_{R^\prime}$ and $Q_\Gamma$ are the amplitudes of the $R_5^-$ and $\Gamma_4^-$ modes, respectively.

\begin{table*}[bht]
\scriptsize
\caption{Fitted coefficients for the Landau energy expansions of the $Pbcm$ [Eq. (\ref{eq:Pbcm})], $Pbnm$ [Eq. (\ref{eq:Pbnm})] and $R3c$ [Eq. (\ref{eq:R3c})] phases of ANO at fixed cubic lattice parameters.
The coefficients are fitted using the units of energy in meV/f.u. and mode amplitudes in \AA.
The values in brackets are instead the coefficients fitted using relative mode amplitudes (i.e. with mode amplitudes normalized to 1 in the relaxed structure of each phase), which also correspond to the energy contributions (meV/f.u) of the related terms to the energy lowering of each phase. The mode amplitudes in each phase can be easily derived from the ratio between the coefficients and the corresponding energy contribution in brackets. 
}
\label{tab:fit}
\begin{ruledtabular}
\begin{tabular}{cllll}
Phase  & \multicolumn{4}{l}{Fitted coefficients (Energy contributions)} \\
\colrule
$Pbcm$ & $A_R          =  -670.5$ ($-113.9$) & $B_R          =   +976.3$ ($ +28.2$) & $C_{R T \Delta}        =  -691.8$ ($-34.3$) & $D_{R T}             = +1721.4$ ($+85.3$) \\
       & $A_T          =  -682.6$ ($-199.2$) & $B_T          =  +1099.1$ ($ +93.6$) & $C_{T \Delta M^\prime} = -4692.4$ ($-55.5$) & $D_{R \Delta}        = -1257.0$ ($-10.6$) \\
       & $A_\Delta     =  -443.3$ ($ -22.0$) & $B_\Delta     = +21447.7$ ($ +53.0$) &                                             & $D_{T \Delta}        = -1959.4$ ($-28.4$) \\
       & $A_{M^\prime} = +2878.0$ ($ +27.7$) \\             
\colrule                                                      
$Pbnm$ & $A_R          =  -670.5$ ($-137.4$) & $B_R          =   +976.3$ ($ +41.0$) & $C_{R M X}             = -1443.2$ ($-17.3$) & $D_{R M}             = +1859.9$ ($+63.1$) \\
       & $A_M          =  -717.2$ ($-118.8$) & $B_M          =  +1151.4$ ($ +31.6$) &                                             & $D_{R X}             = +1079.9$ ($ +0.9$) \\
       & $A_X          = +1547.3$ ($  +6.6$) &                                      &                                             & $D_{M X}             = +1654.3$ ($ +1.2$) \\
\colrule
$R3c$  & $A_{R^\prime} =  -670.5$ ($-299.4$) & $B_{R^\prime} =   +901.3$ ($+179.7$) &                                             & $D_{R^\prime \Gamma} = -2194.2$ ($-60.1$) \\
       & $A_\Gamma     =  -768.8$ ($ -47.1$) & $B_\Gamma     = +14264.8$ ($ +53.6$) \\
\end{tabular}
\end{ruledtabular}
\end{table*}

\begin{figure*}[thb]
\centering
\includegraphics[scale=0.5]{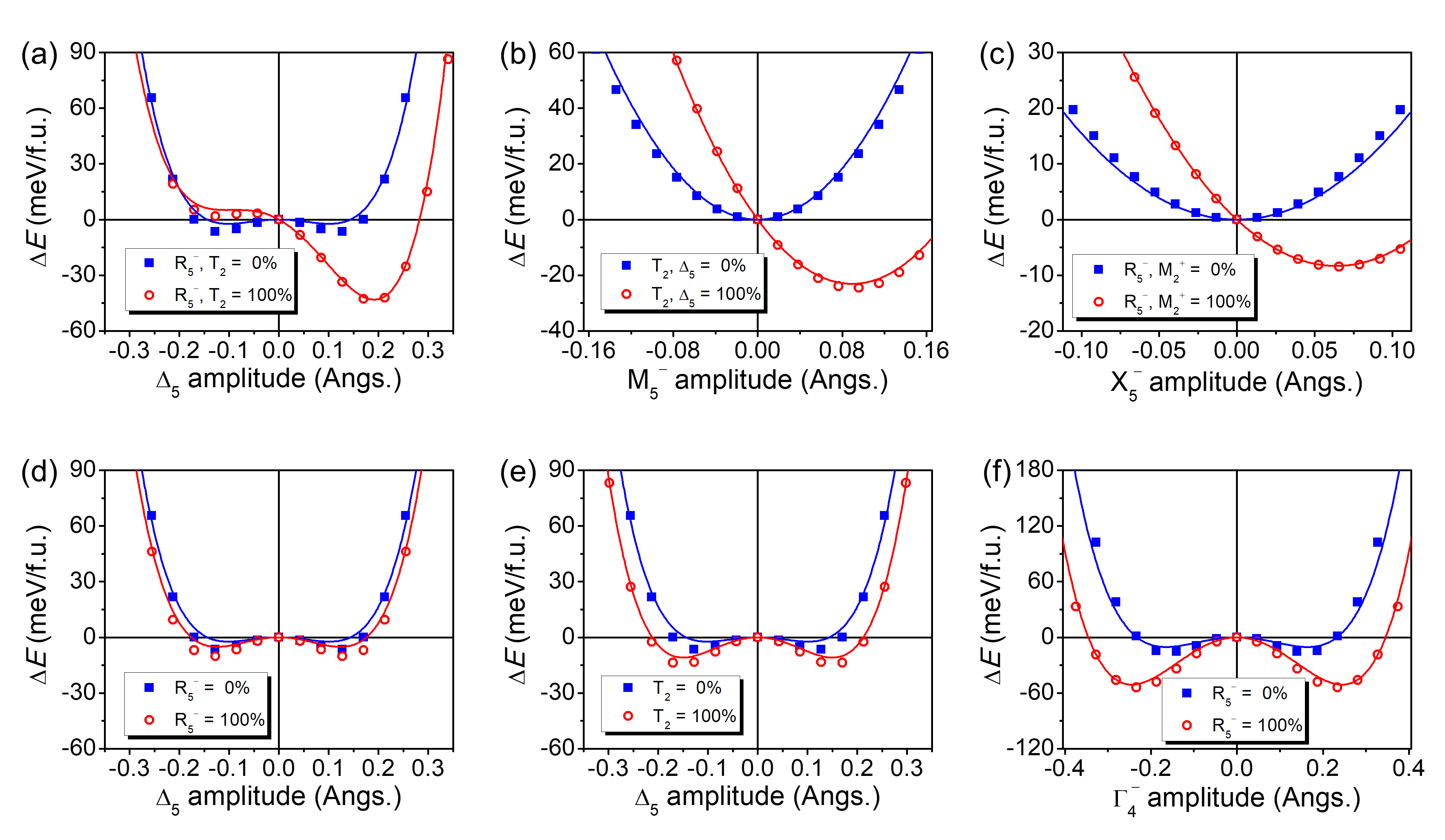}
\caption{Evolution of the energy of ANO when condensing different symmetry adapted mode in the cubic phase (taken as energy reference) in the absence or eventually in the presence of other modes as indicated by the legends.
Mode amplitudes of 100\% correspond to the amplitude in the relaxed (a, b, d, e) $Pbcm$, (c) $Pnma$ and (f) $R3c$ phases with cubic lattice parameters.}
\label{fig:couplings}
\end{figure*}

To quantify the coefficients in the energy expansions, we constructed first-principles data sets for each of the $Pbcm$, $Pbnm$ and $R3c$ phases by introducing relevant lattice-distortion modes with different mode amplitudes to the prototypical $Pm\bar3m$ cubic phase at fixed cubic lattice parameters, and performed the global least-squares fittings of all the energy data points. 
For all the three phases, the fitting shows good quality ($R^2 = 0.9975$), indicating that these energy expressions describe well the associated potential energy surfaces. 
The obtained coefficients are summarized in Table \ref{tab:fit}.

It is worth noticing that the energy expansion of the $Pbcm$ phase [Eq. (\ref{eq:Pbcm})] includes two trilinear coupling terms, i.e. $C_{R T \Delta} Q_R Q_T Q_\Delta$ and $C_{T \Delta M^\prime} Q_T Q_\Delta Q_{M^\prime}$.
We plotted in Figs. \ref{fig:couplings}(a, b) the associated energy curves.
In Fig. \ref{fig:couplings}(a), it is seen that the antipolar $\Delta_5$ mode, when condensed alone in the cubic phase, can only give rise to a shallow and symmetric double well. 
However, combined with the $R_5^-$ and $T_2$ modes, the energy curve becomes highly asymmetric, highlighting the effect of the trilinear coupling term $C_{R T \Delta} Q_R Q_T Q_\Delta$.
In Fig. \ref{fig:couplings}(b), the potential energy curve associated with the $M_5^-$ mode is initially symmetric and single-well shaped when $T_2$ and $\Delta_5$ modes are zero.
However, when the amplitudes of $T_2$ and $\Delta_5$ modes are non-zero, the potential well becomes asymmetric with the bottom of the well shifting to a non-zero amplitude of $M_5^-$ mode.
This is a consequence of the trilinear coupling term $C_{T \Delta M^\prime} Q_T Q_\Delta Q_{M^\prime}$.

In the $Pbnm$ phase, there exists a similar trilinear term $C_{R M X} Q_R Q_M Q_X$ [Eq. (\ref{eq:Pbnm})], which represents the coupling of octahedra rotations $R_5^-$, $M_2^+$, and cation antipolar motion $X_5^-$.
This trilinear coupling has been well-known in literature, as the $Pbnm$ symmetry is the most commonly observed in perovskites \cite{RN230}. 
Fig. \ref{fig:couplings}(c) shows the associated energy curves in ANO. 
By itself, the $X_5^-$ mode is characterized by a single energy well.
The introduction of $R_5^-$ and $M_2^+$ modes leads to the asymmetric shift of the single well due to the trilinear coupling.

Having a closer look at the $Pbcm$ and $Pbnm$ phases of ANO, one can see similarities between the modes involved in the respective trilinear couplings of $R_5^-$, $T_2$ and $\Delta_5$, and of $R_5^-$, $M_2^+$ and $X_5^-$.
The first mode, $R_5^-$, comprises the same $a^-a^-c^0$ rotations present in both phases; because of this we fixed the coefficients $A_R$ and $B_R$ to have the same value in the energy expansions for the two phases [Eqs. (\ref{eq:Pbcm}, \ref{eq:Pbnm})].
The second mode, $T_2$ in $Pbcm$ and $M_2^+$ in $Pbnm$, essentially comprises oxygen octahedra rotations around the $c$-axis, the main difference being the relative signs of the rotations.
In the $M_2^+$ mode, all oxygen octahedra in a $c$-directed column have the same rotation sign, while in $T_2$ the octahedra in a $c$-directed column form a ${+}{+}{-}{-}$ rotation pattern.
The $M_2^+$ and $T_2$ modes having the same mode amplitude show the same absolute rotation angle and have very similar energetics: indeed, as one can see in Tab.~\ref{tab:fit}, $A_T$ and $A_M$, as well as $B_T$ and $B_M$ have almost identical values, lying within 5\% from each other.

The main difference in the energetics comes from the third mode, which is $\Delta_5$ in $Pbcm$ and $X_5^-$ in $Pbnm$.
At a first glance, from the point of view of crystalline structure, these modes are very similar: they both consist of $b$-directed antipolar cation motions manifesting in the same kind of displacement of the Ag cations lying within the same plane normal to the $c$-axis.
In the $X_5^-$ mode, the Ag planes form a ${+}{-}$ displacement pattern, while in the $\Delta_5$ mode they form a ${+}{0}{-}{0}$ pattern.
Interestingly, in the $Pbcm$ phase, the non-zero Ag displacement planes of the $\Delta_5$ mode are centered between the \emph{in-phase} layers of the oxygen octahedral ${+}{+}{-}{-}$ rotation pattern of the $T_2$ mode [Fig.~\ref{fig:struct}(a)], locally reminiscing the distortions observed in the $Pbnm$ phase.
This points towards similar origins for the trilinear coupling of $R_5^-$, $T_2$ and $\Delta_5$ that we report here and the well-known trilinear coupling of $R_5^-$, $M_2^+$ and $X_5^-$.
Despite this similarity, the energy lowering from $I4/mcm$ to $Pbcm$ phase is significantly larger than from $P4/mbm$ to $Pbnm$ [Fig. \ref{fig:phases}(a)].
One reason for this is that inducing the $X_5^-$ mode by itself costs energy, consistently with the positive sign of its quadratic coefficient in the energy expansion ($A_X>0$) and by the single-well nature of the related energy in Fig. \ref{fig:couplings}(c).
On the contrary, the $\Delta_5$ mode is, by itself, unstable in the cubic phase ($A_\Delta<0$), as apparent from the double-well energy profile in Fig. \ref{fig:couplings}(a), allowing for larger trilinear-coupling-mediated energy gains.

In addition to the trilinear couplings, the energy expansion of the $Pbcm$ phase [Eq. (\ref{eq:Pbcm})] also contains three biquadratic coupling terms: $D_{R T} Q_R^2 Q_T^2$, $D_{R \Delta} Q_R^2 Q_\Delta^2$ and $D_{T \Delta} Q_T^2 Q_\Delta^2$. 
Among them, the latter two have negative coefficients, i.e. $D_{R \Delta} < 0$ and $D_{T \Delta} < 0$ (Table \ref{tab:fit}), revealing the cooperative nature of the biquadratic interactions between $R_5^-$ and $\Delta_5$ [Fig. \ref{fig:couplings}(d)] and between $T_2$ and $\Delta_5$ [Fig. \ref{fig:couplings}(e)]. 
In contrast, in $Pbnm$ phase all biquadratic couplings are competitive, with $D_{RM}$, $D_{RX}$, $D_{MX}>0$, which makes another important contribution to the different energy lowering and is ultimately favoring $Pbcm$ over $Pbnm$ in ANO.
Interestingly, we also observe a similar cooperative biquadratic coupling $D_{R^\prime \Gamma} Q_{R^\prime}^2 Q_\Gamma^2$ in the $R3c$ phase [Eq. (\ref{eq:R3c}), Fig. \ref{fig:couplings}(f)].
This cooperative biquadratic coupling in $R3c$ leads to an even more significant energy lowering than the cooperative biquadratic coupling in the $Pbcm$ phase (Table \ref{tab:fit}).

\begin{figure*}[htb]
\centering
\includegraphics[scale=0.5]{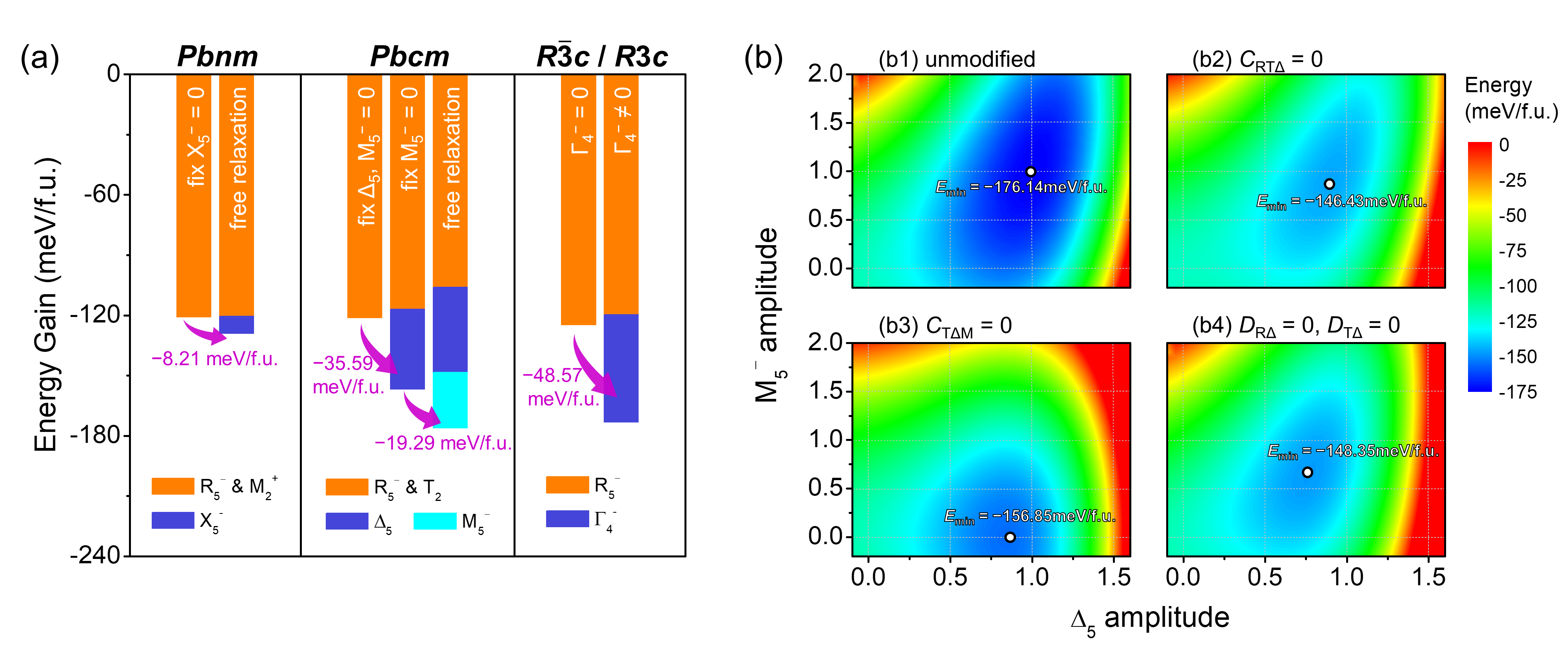}
\caption{(a) Energy lowerings of $Pbcm$, $Pbnm$ and $R\bar{3}c/R3c$ phases with respect to the cubic parent phase when different distortion modes are progressively introduced and relaxed.
(b) Effects of “switching off” some specific couplings on the potential energy surfaces of the $Pbcm$ phase in terms of the amplitudes of $\Delta_5$ and $M_5^-$ modes.
At each point on the plots in (b), the energy value were calculated by relaxing $R_5^-$ and $T_2$ modes at the fixed $\Delta_5$ and $M_5^-$ amplitudes.}
\label{fig:pes}
\end{figure*}

To gain deeper insights on the effects of distinct couplings on the stabilization of the $Pbcm$ phase and to make the above discussion of the couplings more quantitative, we performed Landau-based calculations where different distortion modes are progressively introduced and relaxed.
As shown in Fig. \ref{fig:pes}(a), when only the octahedra rotation modes are allowed to relax while the polar/antipolar modes are fixed to zero, the $Pbnm$, $Pbcm$, and $R\bar3c$ structures exhibit very similar energies, namely $\rm -120.9\ meV/f.u.$, $\rm -121.3\ meV/f.u.$, and $\rm -124.7\ meV/f.u.$, respectively. 
In this case, the $Pbcm$ phase does not show any significant energy advantage over the $Pbnm$ and $R\bar3c$ phases. 

When we further allow the polar/antipolar modes to relax, we find that the $Pbcm$ phase acquires a significant energy lowering. 
As shown in Fig. \ref{fig:pes}(a), by introducing antipolar $\Delta_5$ and $M_5^-$ modes, the energy of $Pbcm$ phase is lowered by 35.59 $\rm meV/f.u.$ and 19.29 $\rm meV/f.u.$, respectively. 
In comparison, the introduction of antipolar $X_5^-$ mode in the $Pbnm$ phase only lowers the energy by 8.21 $\rm meV/f.u.$.
In the $R\bar{3}c$ phase, the condensation of the polar $\Gamma_4^-$ mode also leads to a significant energy reduction of 48.57 $\rm meV/f.u.$, but the resulting $R3c$ phase has an energy still slightly higher than that of the $Pbcm$ phase. 
These results demonstrate the crucial role of the antipolar $\Delta_5$ and $M_5^-$ modes and of their related couplings in stabilizing the $Pbcm$ phase.

\begin{table*}[htb]
\caption{Energy and mode amplitudes of the relaxed $Pbcm$ phase when artificially ``switching off" some specific coupling terms.
The mode amplitudes are normalized to 1 in the real $Pbcm$ phase (``unmodified'' case).}
\label{tab:e_min}
\begin{ruledtabular}
\begin{tabular}{cccccc}
       & $E_{\rm min}$ (meV/f.u.) &  $R_5^-$  &  $T_2$  &  $\Delta_5$  &  $M_5^-$ \\
\colrule
unmodified                  & $-176.14$ & $1.000$ & $1.000$ & $1.000$ & $1.000$ \\
$C_{R T \Delta} = 0$        & $-146.43$ & $0.791$ & $1.011$ & $0.901$ & $0.911$ \\
$C_{T \Delta M^\prime} = 0$ & $-156.85$ & $1.143$ & $0.826$ & $0.839$ & $0.000$ \\
$D_{R \Delta} = 0$          & $-167.29$ & $0.862$ & $1.038$ & $0.964$ & $1.001$ \\
$D_{T \Delta} = 0$          & $-156.53$ & $1.145$ & $0.823$ & $0.836$ & $0.688$ \\
\makecell[c]{$C_{R T \Delta} = 0$, $C_{T \Delta M^\prime} = 0$ } & $-133.19$ & $1.122$ & $0.751$ & $0.697$ & $0.000$ \\
\makecell[c]{$D_{R \Delta}   = 0$, $D_{T \Delta}          = 0$ } & $-148.35$ & $1.047$ & $0.859$ & $0.769$ & $0.660$ \\
\makecell[c]{$C_{R T \Delta} = 0$, $D_{R \Delta}          = 0$, $D_{T \Delta} = 0$ } & $-127.80$ & $1.025$ & $0.800$ & $0.613$ & $0.490$ \\
\makecell[c]{$C_{R T \Delta} = 0$, $C_{T \Delta M^\prime} = 0$, $D_{R \Delta} = 0$, $D_{T \Delta} = 0$ } & $-123.56$ & $1.151$ & $0.679$ & $0.456$ & $0.000$ \\
\end{tabular}
\end{ruledtabular}
\end{table*}

Having the Landau energy expansions, it is also interesting to investigate the effects of ``switching off" some specific couplings by setting their coefficients zero. 
Fig. \ref{fig:pes}(b) presents the energy surfaces with respect to the amplitude of $\Delta_5$ and $M_5^-$ modes from the unmodified Landau energy expansion as well as from some examples of modified versions in which some couplings are artificially ``switched off". 
At each point on the energy surface plots, the $R_5^-$ and $T_2$ modes have been relaxed at the given amplitudes of $\Delta_5$ and $M_5^-$ modes.
We also list in Table \ref{tab:e_min} the energy values and the relative mode amplitudes corresponding to the minimum points of the energy surfaces.

Several interesting features can be seen from Table \ref{tab:e_min} and Fig. \ref{fig:pes}(b).
First, ``switching off" any of the trilinear couplings, i.e. $C_{R T \Delta}$ and $C_{T \Delta M^\prime}$, or the cooperative biquadratic couplings, i.e. $D_{R \Delta}$ and $D_{T \Delta}$, reduces the amplitude of the antipolar $\Delta_5$ mode as compared to the unmodified case.
This is because all these couplings are associated with the antipolar $\Delta_5$ mode and they are cooperative in nature.
Second, when the trilinear coupling term $C_{T \Delta M^\prime}$ is set to zero, the amplitude of the antipolar $M_5^-$ mode will be zero.
This is because the $M_5^-$ mode, by itself, is stable in the cubic phase ($A_{M^\prime} > 0$), and its condensation in the $Pbcm$ phase is due entirely to the trilinear coupling $C_{T \Delta M^\prime}$.
In addition, from Table \ref{tab:e_min}, we can see that the $R_5^-$ and $T_2$ modes always behave in opposite manners, i.e. if one of these two is suppressed, the other one is always enhanced.
This is due to the fact that the coupling between the $R_5^-$ and $T_2$ modes is competitive ($D_{R T} > 0$).
Finally, if any of the cooperative coupling terms is ``switched off", the energy of the $Pbcm$ phase (the minimum of the energy surface) will be increased.
This means that the $Pbcm$ phase would no more be so low in energy and win over in the competition with other polymorphs such as the $R3c$.
All these results demonstrate the crucial role of these cooperative couplings in stabilizing the $Pbcm$ antiferroelectric state.

In summary, we have discussed the characteristics of the Born-Oppenheimer potential energy surface at zero kelvin, explaining why the $Pbcm$ phase, which is a very unusual occurrence in perovskites, exhibits such low energy in ANO. Further research will take temperature-related thermodynamic and kinetic effects into account, so as to better understand how ANO undergoes a complex polymorph evolution process and ultimately stabilizes in the $Pbcm$ phase.

\section{Comparisons between ANO and PZO}

We have analyzed the mode couplings in ANO, highlighting the crucial role of trilinear and cooperative biquadratic couplings of octahedra rotations and cation antipolar motions in stabilizing the $Pbcm$ antiferroelectric state. In the following, we would like to discuss the similarities and differences between ANO and the prototypical antiferroelectric material PZO in the stabilization mechanisms of their respective $Pbcm$ and $Pbam$ antiferroelectric states. 
Previously, J. \'{I}\~{n}iguez, et al. \cite{Iniguez2014} have provided a detailed elucidation of the mode couplings and the stabilization mechanism for the $Pbam$ antiferroelectric state of PZO.
Here, to keep consistency and achieve quantitative comparisons, we briefly reanalyze PZO using the same methodology that we employed for ANO.
It should be noted that our results are consistent with the previous literature  \cite{Iniguez2014}, despite minor differences in methodology.

\begin{table*}[hbt]
\caption{Symmetry-adapted lattice-distortion mode decomposition of the $Pbam$ phase of PZO. 
The structure is obtained by relaxing fully the atomic positions and lattice parameters using GGA-PBEsol functional.
The mode amplitudes are reported in the so-called ``parent-cell-normalized" values of ISODISTORT.}
\label{tab:modePZO}
\begin{ruledtabular}
\begin{tabular}{cllllll}
 & $R_5^-$   & $\Sigma_2$ & $S_2$     & $R_4^-$   &   $M_5^-$ & $X_1^+$   \\
\colrule                                                                          
Amplitude (\AA) & $0.5323$ & $0.4316$  & $0.1366$ & $0.0307$ & $0.0180$ & $0.0179$ \\
Percentage (\%) & $57.91$   & $37.97$    & $3.81$    & $0.19$    & $0.07$    & $0.07$    \\
\end{tabular}
\end{ruledtabular}
\end{table*}

\begin{figure*}[htb]
\centering
\includegraphics[scale=0.45]{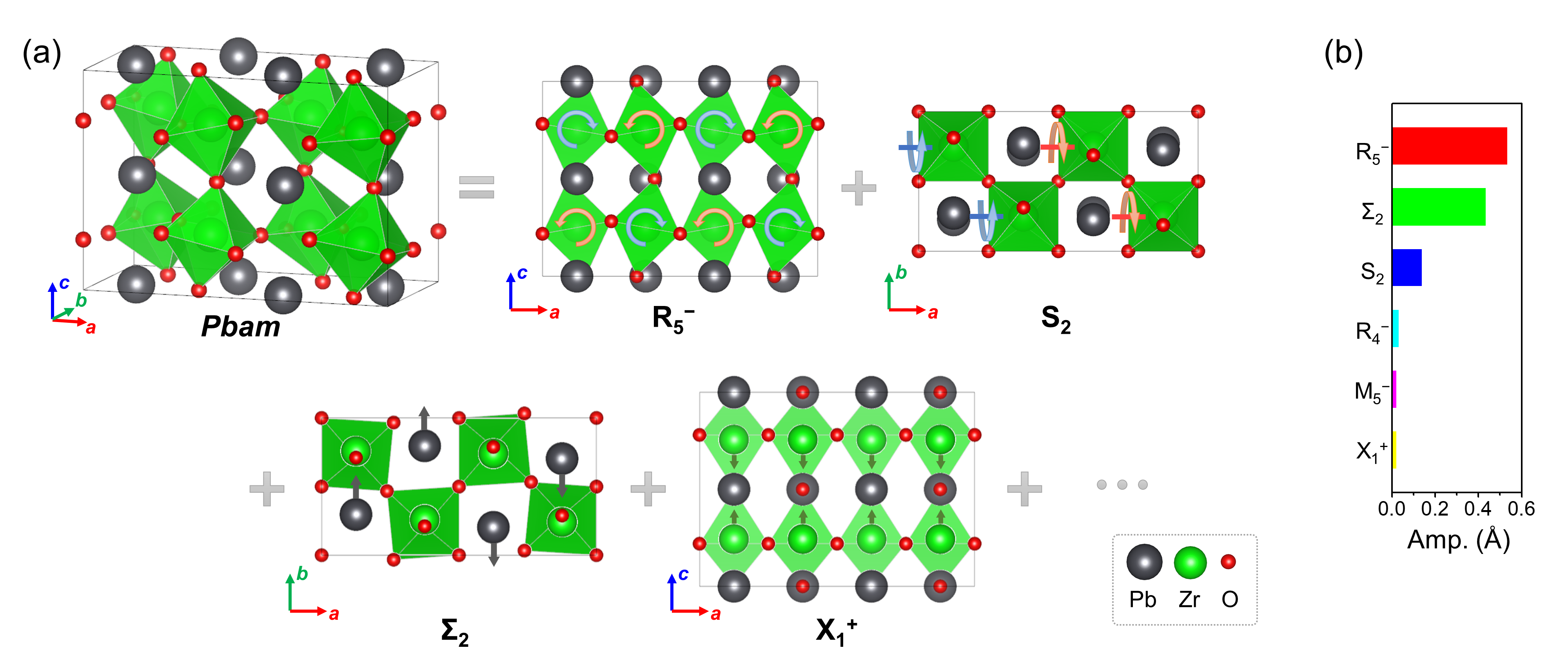}
\caption{
Structure and symmetry-adapted lattice-distortion modes (respect to the cubic reference structure) of PZO in the $Pbam$ phase. (a) Schematics of the $Pbam$ crystal structure and of the lattice-distortion modes of the cubic reference structure contained in the $Pbam$ phase. (b) Mode amplitudes in the $Pbam$ structure, respect to the cubic reference, as optimized with the GGA-PBEsol functional.
}
\label{fig:Pbam_modes}
\end{figure*}

\begin{figure*}[htb]
\centering
\includegraphics[scale=0.5]{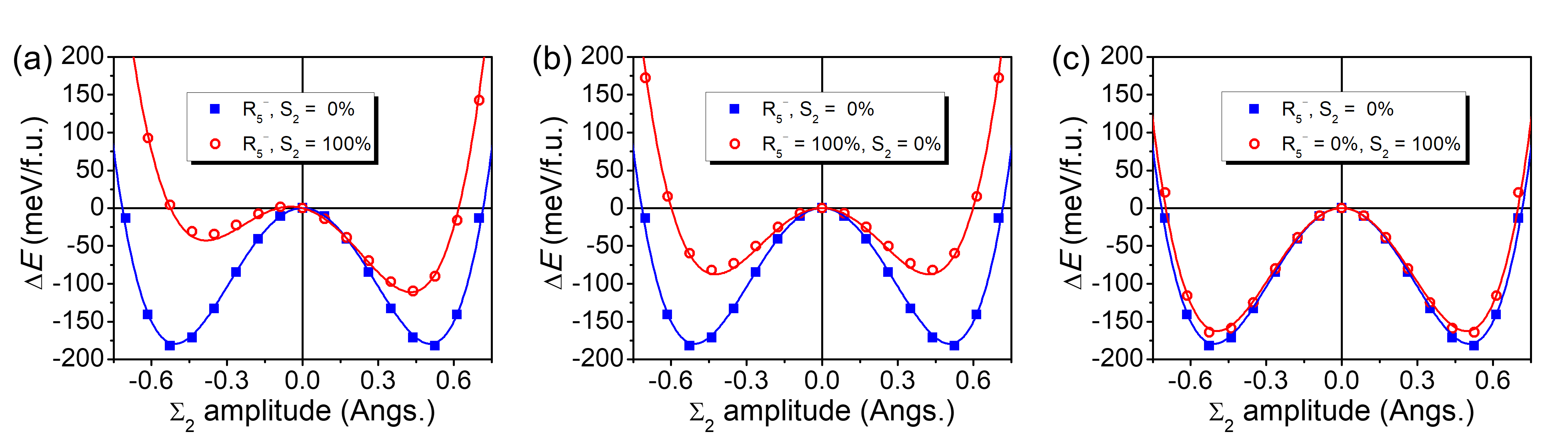}
\caption{
Evolution of the energy of PZO when condensing different symmetry adapted mode in the cubic phase (taken as energy reference) in the absence or eventually in the presence of other modes as indicated by the legends.
Mode amplitudes of 100\% correspond to the relaxed amplitude in the relaxed $Pbam$ phase with cubic lattice parameters.
}
\label{fig:PZO_Pbam}
\end{figure*}

The room-temperature antiferroelectric state of PZO is the $Pbam$ phase.
The dominant symmetry-adapted lattice-distortion modes in the $Pbam$ phase of PZO with respect to its cubic parent phase are [Fig. \ref{fig:Pbam_modes}(a)]: (1) $R_5^-$ mode, the $a^-a^-c^0$ octahedra rotations, (2) $\Sigma_2$ mode, primarily the antipolar Pb motions $\uparrow \uparrow \downarrow \downarrow$ and (3) a $S_2$ mode.
These three modes contribute 57.91\%, 37.97\%, and 3.81\% to the total distortion, respectively [Fig. \ref{fig:Pbam_modes}(b) and Table \ref{tab:modePZO}].
The $S_2$ mode has a complex distortion pattern, whose physical interpretation as a modulated oxygen octahedra rotations was recently given by Shapovalov and Stengel \cite{ShapovalovandStengel}.
The Landau energy expansion of the $Pbam$ phase of PZO can be written as
\begin{equation}
\begin{split}
E_{Pbam} =
& E_{\rm cubic}
+ A_R Q_R^2 + B_R Q_R^4  
+ A_S Q_S^2 + B_S Q_S^4
+ A_\Sigma Q_\Sigma^2 + B_\Sigma Q_\Sigma^4 \\
& + C_{R S \Sigma} Q_R Q_S Q_\Sigma
+ D_{R S} Q_R^2 Q_S^2
+ D_{R \Sigma} Q_R^2 Q_\Sigma^2
+ D_{S \Sigma} Q_S^2 Q_\Sigma^2, 
\end{split}
\label{eq:Pbam}
\end{equation}
where $Q_R$, $Q_S$, and $Q_\Sigma$ are the amplitudes of the $R_5^-$, $S_2$, and $\Sigma_2$ modes, respectively.
Notably, this energy expansion contains a trilinear coupling among the $R_5^-$, $S_2$ and $\Sigma_2$ modes, and biquadratic couplings between any two of the three modes.
It should be kept in mind that the energy expression has been restricted to the three most dominant modes; there are several other modes in the $Pbam$ phase [Fig. \ref{fig:Pbam_modes}(b) and Table \ref{tab:modePZO}], but these modes have been omitted due to their minor amplitudes and negligible energy contributions.
We qualified the coefficient values of Eq. (\ref{eq:Pbam}) by performing a least-squares fitting on a first-principles data set ($R^2 = 0.9952$).
The fitted coefficients are summarized in Table \ref{tab:fitPZO}.
We also show in Fig. \ref{fig:PZO_Pbam} the energy curves associated with the trilinear and biquadratic couplings of the $Pbam$ phase of PZO.

\begin{table*}[ht]
\scriptsize
\caption{Fitted coefficients of the Landau energy expansion of the $Pbam$ phase of PZO [Eq. (\ref{eq:Pbam})] at fixed cubic lattice parameters.
The coefficients are fitted using the units of energy in meV/f.u. and mode amplitudes in \AA.
The values in brackets are instead the coefficients fitted using relative mode amplitudes (i.e. with mode amplitudes normalized to 1 in the relaxed structure of the $Pbam$ phase), which also correspond to the energy contributions (meV/f.u) of the related terms to the energy lowering of this phase. The mode amplitudes in the phase $Pbam$ phase can be easily derived from the ratio between the coefficients and the corresponding energy contribution in brackets.}
\label{tab:fitPZO}
\begin{ruledtabular}
\begin{tabular}{cllll}
Phase  & \multicolumn{4}{l}{Fitted coefficients (Energy contributions)} \\
\colrule
$Pbam$ & $A_R      =  -922.6$ ($-277.1$) & $B_R      = +1073.4$ ($ +96.9$) & $C_{R S \Sigma} = -1075.5$ ($ -38.7$) & $D_{R \Sigma} = +1607.3$ ($ +84.5$) \\
       & $A_S      =  -574.2$ ($ -14.1$) & $B_S      = +2205.8$ ($  +1.3$) &                                       & $D_{S \Sigma} = +2892.7$ ($ +12.4$) \\
       & $A_\Sigma = -1399.4$ ($-244.9$) & $B_\Sigma = +2730.6$ ($ +83.6$) &                                       & $D_{R S}      = +2480.1$ ($ +18.3$) \\
\end{tabular}
\end{ruledtabular}
\end{table*}

There is a certain similarity between the $Pbcm$ phase of ANO and the $Pbam$ phase of PZO in terms of the mode couplings.
First, we note that the trilinear coupling of $R_5^-$, $T_2$ and $\Delta_5$ in ANO appears to be similar to the trilinear coupling of $R_5^-$, $S_2$ and $\Sigma_2$ in PZO: they involve the same $a^-a^-c^0$ octahedra rotations around $b$-axis in the $R_5^-$ mode, both the $T_2$ and $S_2$ can be interpreted as octahedra rotations around $c$- and $a$-axis, respectively, modulated along the respective rotation axis direction \cite{ShapovalovandStengel}, and both the $\Delta_5$ and $\Sigma_2$ consist of $b$-directed displacements of cations.
Second, there is an additional trilinear coupling of $T_2$, $\Delta_5$ and $M_5^-$ modes identified in ANO, whose counterpart in the $Pbam$ phase of PZO, despite being overlooked by the energy expansion Eq. (\ref{eq:Pbam}), is also allowed by symmetry. 
From symmetry analysis, we identified a trilinear coupling in the $Pbam$ phase of PZO, which involves  $S_2$, $\Sigma_2$ and $X_1^+$ modes, that is similar to the trilinear coupling of $T_2$, $\Delta_5$ and $M_5^-$ in the $Pbcm$ phase of ANO.
The $X_1^+$ mode is also mainly the antipolar motion of cations [Fig. \ref{fig:Pbam_modes}(a)], just like the $M_5^-$ mode in the $Pbcm$ phase of ANO.
However, due to its negligible amplitude and energy contribution, the $X_1^+$ mode is usually not considered when constructing the energy expansion.
Finally, all the biquadratic coupling terms between any two of the modes are allowed by the symmetry for both $Pbcm$ and $Pbam$.
Therefore, from the qualitative perspective, the antiferroelectric $Pbcm$ phase in ANO and $Pbam$ phase in PZO are similar in terms of mode couplings.

Nevertheless, some important differences can be noticed if we inspect more closely the relative mode contributions in the two materials.
Among the three most dominant lattice-distortion modes of ANO, the antipolar $\Delta_5$ mode is remarkably less important than the other two modes, the rotational $R_5^-$ and $T_2$, as its amplitude and energy contribution are notably smaller.
In this sense, the two rotational modes can be considered as the ``primary'' modes defining the symmetry of the $Pbcm$ phase of ANO, and the antipolar $\Delta_5$ mode as the ``secondary'' mode whose amplitude and energy contribution are largely enhanced by the coexistence with the primary modes.
In PZO, the roles of rotational and antipolar modes are switched: there, the ``primary'' modes are rotational $R_5^-$ and antipolar $\Sigma_2$, which have larger amplitudes and give significantly larger contributions to the energetics of the $Pbam$ phase, while the remaining rotational $S_2$ mode is ``secondary''.
Thus, the antipolar cation motions in the antiferroelectric state of PZO appear to be more dominant than in ANO.

The difference between ANO and PZO lies not only in the relative importance of the antipolar modes with respect to the rotational modes, but also in how these modes interact with each other.
The nature of a specific coupling term in the cubic phase, whether it is cooperative or competitive, can be judged by the sign of the respective coefficient in the energy expansion. 
However, it is not always straightforward to know the overall coupling effect between any two modes in the strongly distorted low symmetry phase, since multiple coupling terms may be involved.
Here, we propose an approach that more thoroughly evaluates the combined impact of all relevant coupling terms, thereby ascertaining the overall coupling effect between any two modes. 
The detailed derivations are presented in the Appendix. 
Adopting this approach, we found that in ANO the overall coupling effect of the octahedra rotational modes, i.e. $R_5^-$ and $T_2$, on the antipolar $\Delta_5$ mode remain essentially cooperative in the $Pbcm$ phase (calculated based on the energy coefficients listed in Table \ref{tab:fit} and the mode amplitudes $Q_R = 0.4122$ \AA, $Q_T = 0.5402$ \AA, $Q_\Delta = 0.2230$ \AA, $Q_{M^\prime} = 0.0982$ \AA):
\begin{equation}
\begin{split}
\frac{\partial Q_\Delta}{\partial Q_R} \bigg|_{Pbcm} &= 
-\frac{C_{R T \Delta}Q_T + 4D_{R \Delta}Q_RQ_\Delta}
{2A_\Delta + 12B_\Delta Q_\Delta^2 + 2D_{R \Delta}Q_R^2 + 2D_{T \Delta}Q_T^2}
=+0.081 > 0, \\
\frac{\partial Q_\Delta}{\partial Q_T} \bigg|_{Pbcm} &= 
-\frac{C_{R T \Delta}Q_R + C_{T \Delta M^\prime}Q_{M^\prime} + 4D_{T \Delta}Q_TQ_\Delta}
{2A_\Delta + 12B_\Delta Q_\Delta^2 + 2D_{R \Delta}Q_R^2 + 2D_{T \Delta}Q_T^2}
=+0.163 > 0.
\end{split} 
\label{eq:dDT}
\end{equation}
This can also be evidenced from Figs. \ref{fig:couplings}(a, d, e), where both the first-principles calculation and the energy expansion-based simulation show that the amplitude of the $\Delta_5$ mode is largely enhanced when coexisting with $R_5^-$ or/and $T_2$ modes. 
Instead for PZO, by the same analysis, we find that the overall coupling effect of $R_5^-$ and $S_2$ modes on the antipolar $\Sigma_2$ mode remain essentially competitive in the $Pbam$ phase (calculated based on the energy coefficients listed in Table \ref{tab:fitPZO} and the mode amplitudes $Q_R = 0.5481$ \AA, $Q_S = 0.1568$ \AA, $Q_\Sigma = 0.4184$ \AA):
\begin{equation} 
\begin{split}
\frac{\partial Q_\Sigma}{\partial Q_R} \bigg|_{Pbam} &= 
-\frac{C_{R S \Sigma}Q_S + 4D_{R \Sigma}Q_RQ_\Sigma}
{2A_\Sigma + 12B_\Sigma Q_\Sigma^2 + 2D_{R \Sigma}Q_R^2 + 2D_{S \Sigma}Q_S^2}
=-0.323 < 0, \\
\frac{\partial Q_\Sigma}{\partial Q_S} \bigg|_{Pbam} &= 
-\frac{C_{R S \Sigma}Q_R + 4D_{S \Sigma}Q_SQ_\Sigma}
{2A_\Sigma + 12B_\Sigma Q_\Sigma^2 + 2D_{R \Sigma}Q_R^2 + 2D_{S \Sigma}Q_S^2}
=-0.042 < 0.
\end{split} 
\label{eq:dSM}
\end{equation}
As can be seen from Fig. \ref{fig:PZO_Pbam}, coexistence with the $R_5^-$ or/and $S_2$ modes suppresses the amplitude of the $\Sigma_2$ mode.

To summarize, although very similar antipolar and octahedra rotation modes -- trilinearly coupled by symmetry -- are involved in both cases, we find a slightly different story in ANO and PZO.
In ANO, the antipolar $\Delta_5$ mode by itself only shows a weak instability, in contrast with the strong instability of the antipolar $\Sigma_2$ mode in PZO.
The emergence of the $\Delta_5$ mode in the antiferroelectric $Pbcm$ phase of ANO is at first favored by the trilinear coupling with the rotational $R_5^-$ and $T_2$ modes.
This behavior reminds us of the concept of hybrid-improper ferroelectricity \cite{EricNature2008, RN229, RN214}, and suggests a more hybrid-improper-like nature of the antipolar cation motions in ANO \cite{RN577}. 
Then, this $\Delta_5$ mode is also significantly enhanced by cooperative biquadratic couplings with $R_5^-$ and $T_2$ modes, which corresponds to a so-called triggered mechanism \cite{Ghosez_and_Junquera_2022}. In comparison, the $\Sigma_2$ antipolar cation motions in PZO appear more proper and compete with $R_5^-$ and $S_2$ modes. 
This observation highlights the importance of octahedra rotations in inducing the cation antipolar motions in ANO.
Usually, the octahedra rotations can be tuned by various factors such as strain, pressure, doping, defects, etc \cite{RN474, RN725, RN198, RN726}.
In this regard, investigating the control of octahedra rotations so as to modulate the AFE property in ANO will also be an interesting perspective for a future work.

\section{Conclusions}
In this study, we have emphasized the crucial role of trilinear and biquadratic couplings of octahedra rotations and cation antipolar motions in stabilizing the $Pbcm$ phase in ANO.
These anharmonic couplings contribute significantly to lower the energy, making the $Pbcm$ phase energetically more favorable compared to other competing variants.
We have also conducted a comparative discussion between the antiferroelectric phases of ANO and PZO, which are distinct but similarly combine antipolar and octahedra rotation motions trilinearly coupled by symmetry. 
Unlike in the $Pbam$ phase of PZO, the antipolar motions in $Pbcm$ phase of ANO appear to be secondary. 
Their significant amplitude in the $Pbcm$ phase arises from the combination of hybrid-improper (i.e. trilinear coupling with primary octahedra rotations) and triggered (i.e. cooperative biquadratic coupling with primary octahedra rotations) mechanisms. 
This contrasts with PZO in which antipolar motions are primary and compete with octahedra rotations.  

\begin{acknowledgments}
This work was supported by the European Union’s Horizon 2020 research and innovation program under grant agreement No. 964931 (TSAR) and by F.R.S.-FNRS Belgium under PDR grant T.0107.20 (PROMOSPAN). H.Z. acknowledges the International Postdoctoral Exchange Fellowship (PC2020060) and the Research IPD-STEMA Program. S.A. acknowledges Erasmus+International Credit Mobility program 2023 of Université de Liège  and the PRD-CCD ARES 2019-2024 project: {\it Le coltan du Kivu: Capacité de traitement physico-chimique et études d'applications} (Grant No. 32210). The authors acknowledge the use of the CECI supercomputer facilities funded by the F.R.S-FNRS (Grant No. 2.5020.1) and of the Tier-1 supercomputer of the Fédération Wallonie-Bruxelles funded by the Walloon Region (Grant No. 1117545).
\end{acknowledgments}

\appendix

\section*{Appendix: Evaluation of overall coupling effect between distortion modes}

Having the energy expansion and the related coefficients, it is relatively easy to judge from the sign of the coefficients of specific coupling terms which modes are initially competitive or cooperative in the cubic reference.
However, it is not always so straightforward to determine the overall coupling effect between different modes in the low-energy phases, especially when multiple coupling terms are involved.
In this Appendix, we present an approach to comprehensively assess the whole effect of all relevant coupling terms, thereby evaluate the overall impact of one mode on the other. 

Let us consider a typical example in a relatively general sense, where three modes $P$, $Q_1$ and $Q_2$ are mutually coupled and the energy expansion is given by
\begin{equation} 
\begin{split}
E = E_0 + A P^2 + B P^4 
+ C Q_1 Q_2 P + D_1 Q_1^2 P^2 + D_2 Q_2^2 P^2,
\end{split}
\label{eq:EPQ}
\end{equation}
where $A$, $B$, $C$, $D_1$ and $D_2$ are energy coefficients, and $E_0$ includes all the energy contributions irrelevant to $P$.
This energy expansion contains one trilinear coupling term, and two biquadratic terms.
The goal is to reveal how $P$ will be affected if small perturbation is applied on $Q_1$ or $Q_2$ near the energy minimum point.

The $P_0$ (the relaxed value of $P$ at given $Q_1$ and $Q_2$) minimizes the energy and must satisfy the condition $\frac{\partial E}{\partial P} |_{P_0}= 0$, which yields
\begin{equation} 
\begin{split}
2AP_0 + 4BP_0^3 + C Q_1 Q_2 + 2 D_1 Q_1^2 P_0 + 2 D_2 Q_2^2 P_0 = 0.
\end{split}
\label{eq:PQ}
\end{equation}
This equation gives an implicit expression of the function $P_0 = P_0(Q_1, Q_2)$.  
Taking the partial derivatives with respect to $Q_1$ and $Q_2$, respectively, on both side of Eq. (\ref{eq:PQ}), we have

\begin{equation} 
\begin{split}
2A \frac{\partial P_0}{\partial Q_1} + 12BP_0^2 \frac{\partial P_0}{\partial Q_1} + C Q_2 + 4 D_1 Q_1 P_0 + 2 D_1 Q_1^2 \frac{\partial P_0}{\partial Q_1} + 2 D_2 Q_2^2 \frac{\partial P_0}{\partial Q_1} = 0, \\
2A \frac{\partial P_0}{\partial Q_2} + 12BP_0^2 \frac{\partial P_0}{\partial Q_2} + C Q_1 + 2 D_1 Q_1^2 \frac{\partial P_0}{\partial Q_2} + 4 D_2 Q_2 P_0 + 2 D_2 Q_2^2 \frac{\partial P_0}{\partial Q_2} = 0.
\end{split}
\label{eq:PQ1}
\end{equation}

Solving $\frac{\partial P_0}{\partial Q_1}$ and $\frac{\partial P_0}{\partial Q_2}$ from Eq. (\ref{eq:PQ1}), and evaluating at the energy minimum point (denoted as ``min$E$'', where $P_0 = P_{00}$, $Q_1 = Q_{10}$, $Q_2 = Q_{20}$), we can get

\begin{equation} 
\begin{split}
\frac{\partial P_0}{\partial Q_1} \bigg|_{{\rm min} E} & =
- \frac{C Q_{20} + 4 D_1 Q_{10} P_{00}}{2A + 12BP_{00}^2 + 2D_1Q_{10}^2 + 2D_2Q_{20}^2}, \\
\frac{\partial P_0}{\partial Q_2} \bigg|_{{\rm min} E} & =
- \frac{C Q_{10} + 4 D_2 Q_{20} P_{00}}{2A + 12BP_{00}^2 + 2D_1Q_{10}^2 + 2D_2Q_{20}^2}.
\end{split}
\label{eq:partial}
\end{equation}

The partial derivatives of $P_0$ with respect to $Q_1$ and $Q_2$ provide the evaluation of the overall impact of $Q_1$ and $Q_2$ on $P$ at the energy minimum point, respectively, with a positive value meaning a cooperative effect, and negative meaning competitive.

In the specific case of ANO, replacing $P$ by $Q_\Delta$, $Q_1$ by $Q_R$, and $Q_2$ by $Q_T$ and considering an additional term $C_{T \Delta M} Q_T Q_\Delta Q_M$ in the energy expansion of the $Pbcm$ phase of ANO, Eq. (\ref{eq:dDT}) can be derived following the same procedure.

In the specific case of PZO, replacing $P$ by $Q_\Sigma$, $Q_1$ by $Q_R$, and $Q_2$ by $Q_S$, Eq. (\ref{eq:partial}) is directly turned into Eq. (\ref{eq:dSM}). 

As a final remark, we point out that the equations can be further simplified if the mode amplitudes are relatively evaluated (renormalized so as to make $P_{00}' = Q'_{10} = Q'_{20} = 1$ at the energy minimum point):
\begin{equation} 
\begin{split}
\frac{\partial P_0'}{\partial Q'_1} \bigg|_{{\rm min} E} & =
- \frac{C' + 4 D'_1}{2A' + 12B' + 2D'_1 + 2D'_2}, \\
\frac{\partial P_0'}{\partial Q'_2} \bigg|_{{\rm min} E} & =
- \frac{C' + 4 D'_2}{2A' + 12B' + 2D'_1 + 2D'_2},
\end{split}
\label{eq:partial2}
\end{equation}
where the primed coefficients have the physical meaning of their respective energy contributions in the studied phase (corresponding to the values in parentheses in Tabs. \ref{tab:fit}, \ref{tab:fitPZO}).
In cases where only the qualitative nature of the coupling effect is of concern, rather than its quantitative strength, this simplified form is easier to use.
We notice that using such primed equivalent to Eq. (5) and (6) would not change our conclusions.

\bibliography{reference}
\bibliographystyle{unsrt}

\end{document}